\documentclass[runningheads]{llncs}
\usepackage[T1]{fontenc}
\usepackage{graphicx}
\usepackage{graphicx}
\usepackage{subcaption}
\begin{document}
\title{SRWToolkit: An Open Source Wizard of Oz Toolkit to Create Social Robotic Avatars}
\titlerunning{SRWToolkit: An Open Source Wizard of Oz Toolkit}
\author{Atikkhan Faridkhan Nilgar\inst{1,2}\orcidID{0009-0001-4948-614X} \and Kristof Van Laerhoven\inst{1}\orcidID{0000-0001-5296-5347} \and Ayub Kinoti\inst{3}\orcidID{0009-0005-0874-0577}
}
\authorrunning{A. Nilgar et al.}
%
\institute{University of Siegen, 57076 Siegen, Germany\and Honda Research Institute Europe GmbH, 63073 Offenbach am Main, Germany\\
\and Dedan Kimathi University of Technology, Nyeri, Kenya \\ \email{atikkhan.nilgar@uni-siegen.de}}

\maketitle
%
\begin{abstract}
We present SRWToolkit, an open-source Wizard of Oz toolkit designed to facilitate the rapid prototyping of social robotic avatars powered by local large language models (LLMs). Our web-based toolkit enables multimodal interaction through text input, button-activated speech, and wake-word command. The toolkit offers real-time configuration of avatar appearance, behavior, language, and voice via an intuitive control panel. In contrast to prior works that rely on cloud-based LLMs services, SRWToolkit emphasizes modularity and ensures on-device functionality through local LLM inference. In our small-scale user study, [n = 11] participants created and interacted with diverse robotic roles (hospital receptionist, mathematics teacher, and driving assistant), which demonstrated positive outcomes in the toolkit's usability, trust, and user experience. The toolkit enables rapid and efficient development of robot characters customized to researchers’ needs, supporting scalable research in human-robot interaction.

\end{abstract}
\begin{sloppypar}
\section{Introduction and Related Work}
Social robotics research often encounters significant barriers due to the complexity and multidisciplinary nature of building interactive and engaging robot systems from scratch. These challenges require expertise in robotics, artificial intelligence (AI), and human-computer interaction design, limiting rapid experimentation and prototyping \cite{Smit2025Enhancing,Schl_gl_2014_Wizard,marge2017applyingwizardofoztechniquemultimodal,Fang2024ONLLM}. Traditional development approaches can be cumbersome, costly, and time-consuming, preventing greater participation from researchers who lack extensive technical backgrounds \cite{Rietz2021WoZ4U}. 

Current advances in large language models (LLMs), specifically GPT-4, have demonstrated impressive conversational capabilities, significantly enhancing natural language understanding and generation \cite{Fang2024ONLLM}. By embedding these models within a modular Wizard of Oz (WoZ) framework, researchers can rapidly prototype sophisticated human-robot interaction (HRI) scenarios while significantly reducing technical complexity.
Prior studies have extensively explored WoZ methodologies as a way to rapidly prototype interactive systems without extensive programming. For instance, WoZ4U \cite{Rietz2021WoZ4U} is an open-source web-based interface designed to facilitate human-robot interaction studies by allowing experimenters to control robot behaviors in real-time with minimal coding effort. However, WoZ4U primarily addresses manual wizard control and does not integrate automated conversational agents. 

More recently, Fang et al. \cite{Fang2024ONLLM} explored “LLM Wizards”, where LLMs themselves act as automated wizards within WoZ setups. This approach showed a significant reduction in manual workload while enabling richer, more scalable interactions. Such integration allows researchers to benefit from the conversational skills of contemporary LLMs, creating natural, contextually relevant interactions without extensive manual scripting or training. Similarly, the WebWOZ \cite{Schl_gl_2014_Wizard} platform offers a generic architecture aimed specifically at language technology research, providing modular, scalable support for speech and text modalities. Popular LuxAI’s QTrobot 
 (\url{https://luxai.com})
also integrates GPT-based dialogue to support learning in educational and therapeutic contexts. However, most of these WoZ systems rely on cloud-based LLM inference, which raises concerns related to data privacy, user consent regarding data handling and storage, as well as increased latency in real-time interactions.

We address these challenges by presenting a modular WoZ toolkit that integrates local LLMs to enable the rapid development and deployment of social robotic avatars. These avatars refer to virtual representations of physical social robots, typically displayed on screens. Although they lack physical presence, they often retain similar interaction behaviors and characteristics, making them useful in scenarios where physical embodiment is not required. Our toolkit builds on prior works' insights, combining the strengths of WoZ4U’s web-based modularity and LLM Wizards' automation capabilities. We provide a web-based module toolkit that supports multimodal interaction, secure local LLM execution, and comprehensive session-based log-in to facilitate adaptive interaction.
\end{sloppypar}

\section{Technical Architecture}
The social robotic avatar toolkit consists of three interconnected layers: the frontend, the backend, and the LLMs, which are detailed below.

\begin{figure}[]
    \centering
    \begin{minipage}[t]{0.49\textwidth}
        \centering
        \includegraphics[width=\textwidth]{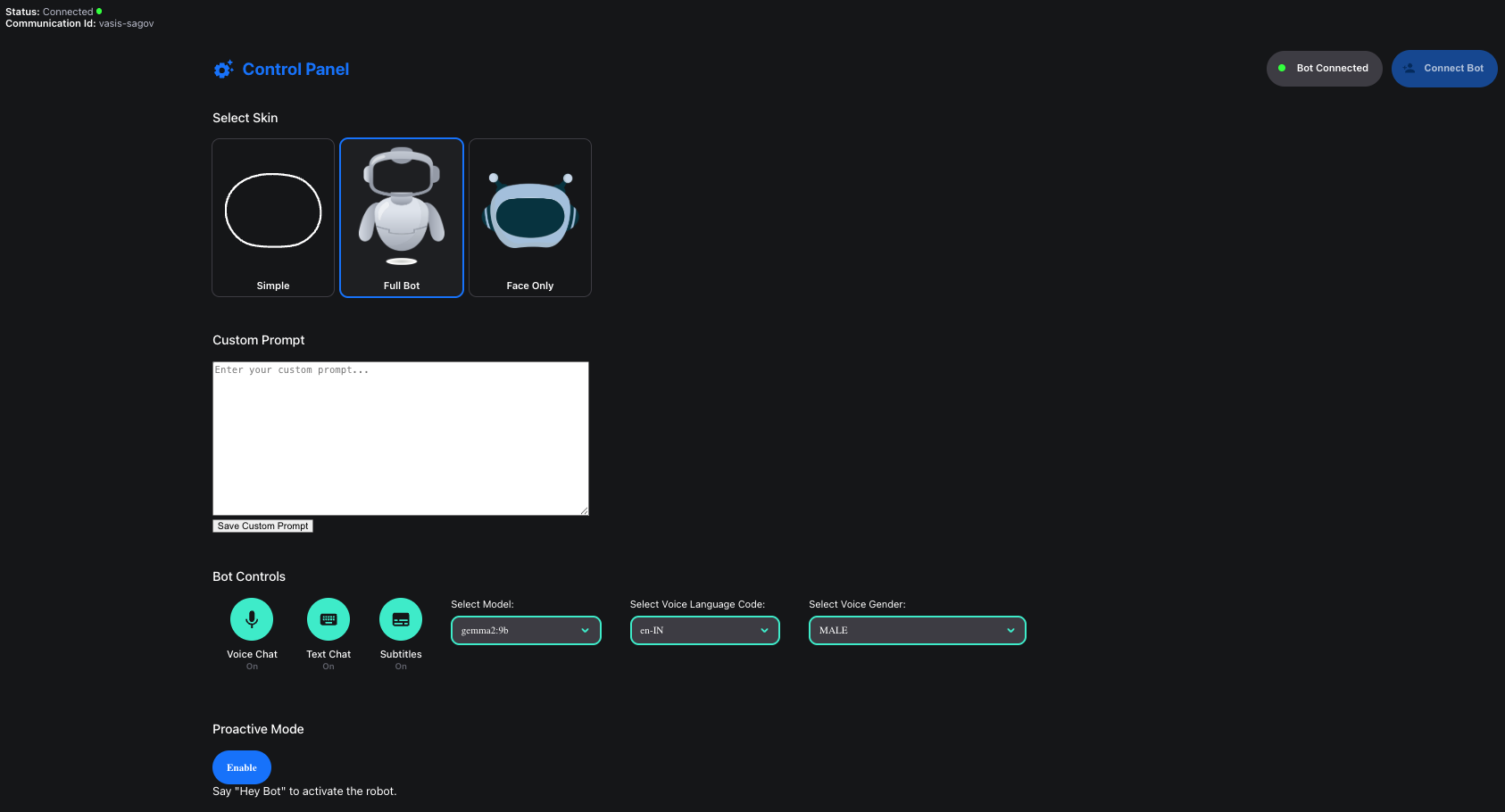}
        \caption{Control Panel allows administrators to manage configurations.}
        \label{fig:control_panel_screen}
    \end{minipage}
    \hfill
    \begin{minipage}[t]{0.49\textwidth}
        \centering
        \includegraphics[width=\textwidth]{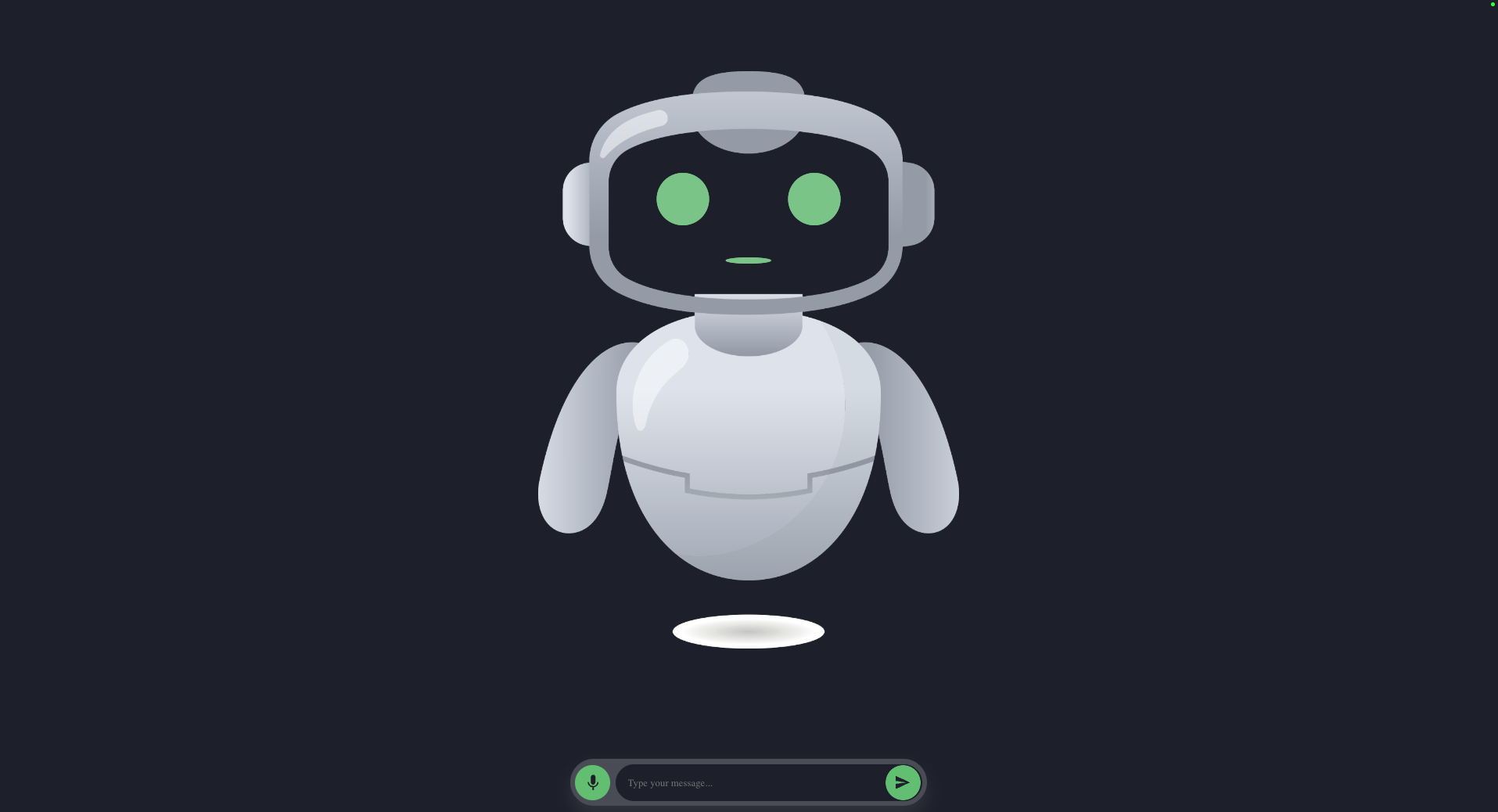}
        \caption{Social Robot Screen for users to interact via text and voice inputs.}
        \label{fig:social_robot_screen}
    \end{minipage}
\end{figure}

\begin{sloppypar}
\textbf{Frontend Implementation.} The frontend is implemented as a browser-based application using JavaScript. The
development environment utilizes node.js (v20.15.0) and yarn (v1.22.22) as
the package manager. The frontend includes two main interfaces:  the Control Panel (Figure \ref{fig:control_panel_screen}) and the Social Robot Screen (Figure  \ref{fig:social_robot_screen}). The control panel is developed using React.js (\url{https://react.dev}) and Redux (\url{https://redux.js.org}). It allows administrators to initiate sessions with human-readable communication IDs, configure avatar visuals, select communication languages, enable or disable interaction modes, choose LLM models, write custom prompts (LLM prompts to create a role/character of the robot), and define the voice gender. These configurations are sent in real time to the backend and stored in MongoDB (\url{https://mongodb.com}). WebSocket (\url{https://dev.mozilla.org/en-US/docs/Web/API/WebSockets\_API}) ensures status updates, including robot state and connectivity, which are sent to the control panel every five seconds. The social robot screen is created using React.js to deliver a dynamic and interactive interface for end-users.

The social robot screen facilitates user interaction through text input, voice input activated by a button, and voice commands triggered by a wake word (“Hey Bot”). Voice inputs are recorded using the MediaRecorder (\url{https://dev.mozilla.org/en-US/docs/Web/API/MediaRecorder}), while the proactive interaction mode uses the SpeechRecognition (\url{https://dev.mozilla.org/en-US/docs/Web/API/SpeechRecognition}) to listen for the phrase of the wake word. Redux is used to manage dynamic avatar states such as animations, blinking, and response indicators. Inputs, such as text or base64-encoded audio, are transmitted via persistent WebSocket connections to the backend. Audio responses from the backend are decoded and played using the browser’s audio.

\textbf{Backend Implementation.} The backend is implemented in Python using FastAPI (\url{https://fastapi.tiangolo.com}). It serves as the central processing unit, which handles input processing, configuration management, data storage, and communication with the LLMs. It operates on-server using Uvicorn (\url{https://www.uvicorn.org}) to support asynchronous interactions. 
Dependencies are managed via Poetry (\url{https://python-poetry.org}). The backend handles WebSocket connections with both the social robot and control panel interfaces and parses incoming data. Voice inputs are transcribed into text using the Google Cloud (\url{https://cloud.google.com}) Speech-to-Text API. The transcribed or original text and the user’s conversation history are forwarded to the LLM for context-aware response generation. The response is converted to speech using the Google Cloud Text-to-Speech API and sent back to the frontend. Data storage is managed using MongoDB, containerized with Docker (\url{https://docker.com}). The database collects ‘communications’ for session metadata and configurations, ‘chat\_messages’ for user input, custom prompts, and LLM response logs. All deployment is restricted to a local network environment to maintain security. The LLM server does not store any audio or conversational content; it retains only anonymized metadata.

\textbf{LLMs Implementation.} We used Ollama (\url{https://ollama.com}), a lightweight framework for deploying and serving large-scale language models via API on the local device (macOS, M2 Pro, Sequoia 15.0.1, 32 GB). We included Llama 3.2, Gemma 2.0, Phi 3.5, Qwen 2.5, and Nemotron-mini 1.0, all of which are pulled and served through Ollama’s API. The LLM API is configured via a one-step port setup that links the toolkit to the GPU of the local device where Ollama runs. This ensures a secure, high-performance, and low-latency environment for model inference. The backend sends HTTP POST requests to the LLM API containing user queries and conversation context. The LLM processes these inputs and returns JSON-formatted responses. None of the data is retained by Ollama, ensuring that all LLM processing remains ephemeral and respects user privacy.
\begin{figure}[]
    \centering
    \includegraphics[width=1\textwidth]{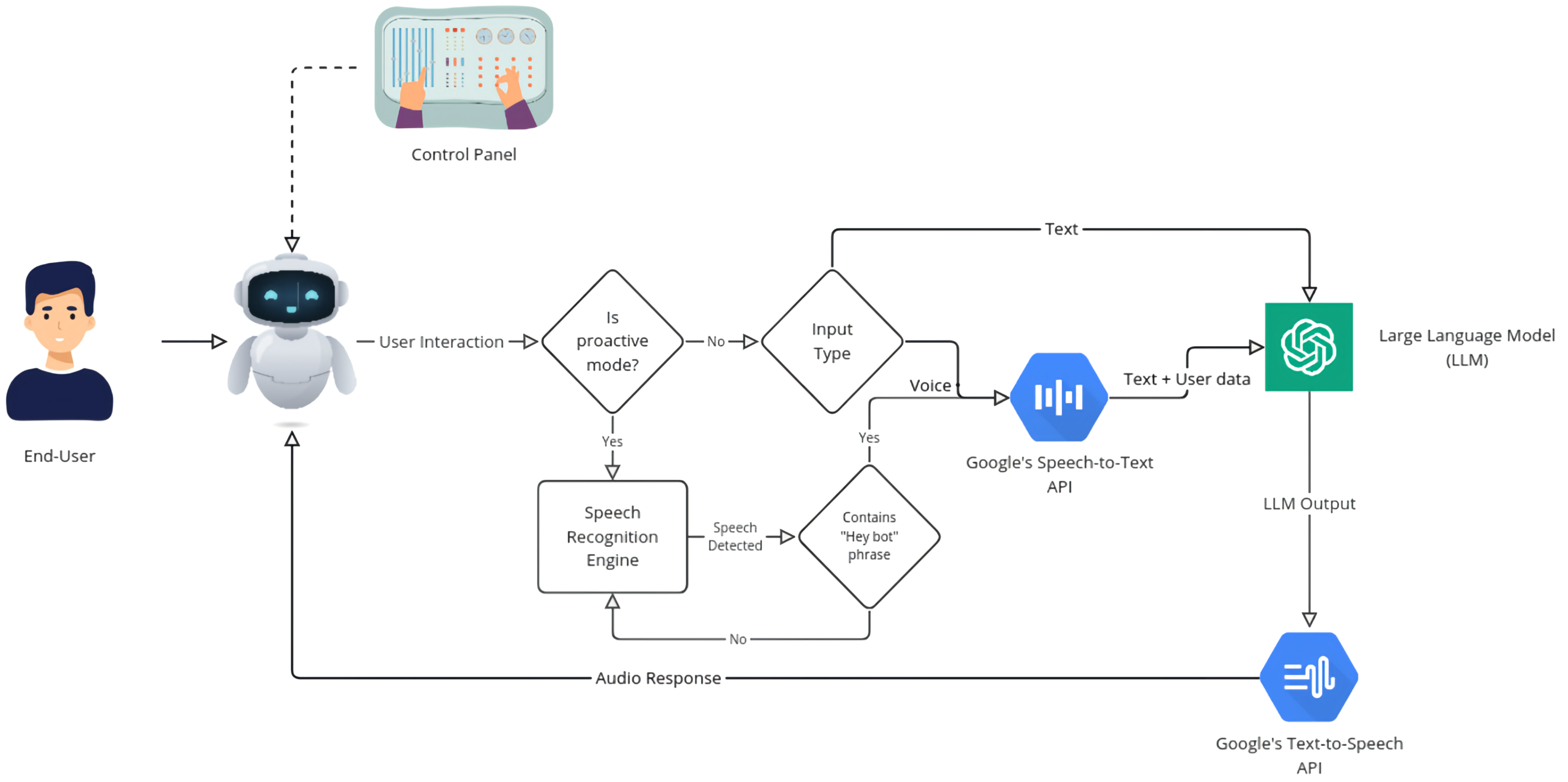}
    \caption{Illustration of the end user interaction process with the social robot screen interface, highlighting the flow of communication and the integration of LLMs for seamless interaction.} 
    \label{fig:Workflow}
\end{figure}

\textbf{Workflow.} Figure~\ref{fig:Workflow} illustrates the interaction flow between the end-user and the social robot, highlighting how the toolkit processes both text and voice inputs to deliver seamless, multimodal communication. Users can interact via text or voice. In proactive mode, the browser’s speech recognition listens for the wake phrase “Hey bot,” activating hands-free interaction. The system continues listening for 5 seconds, enabling dynamic and intuitive engagement.
For non-proactive input, the system detects the input type (text or button activated voice). Text messages are sent directly to the backend, while voice inputs are transcribed using the Google Speech-to-Text API. The resulting text, combined with administrator's custom prompt is processed by the LLM to generate personalized, context-aware responses. The response by LLMs is converted to speech via the Google Text-to-Speech API for natural verbal communication. From the control panel (Figure~\ref{fig:control_panel_screen}) administrators can easily configure interaction modes, select LLM models, write context-based custom prompts, change languages, and adjust voice genders in real-time.
\end{sloppypar}

\begin{sloppypar}
\section{User Study}
\begin{figure*}[h]
    \centering
    \begin{minipage}[c]{0.49\textwidth}
        \centering
        \begin{subfigure}[b]{\textwidth}
            \centering
            \includegraphics[width=\textwidth]{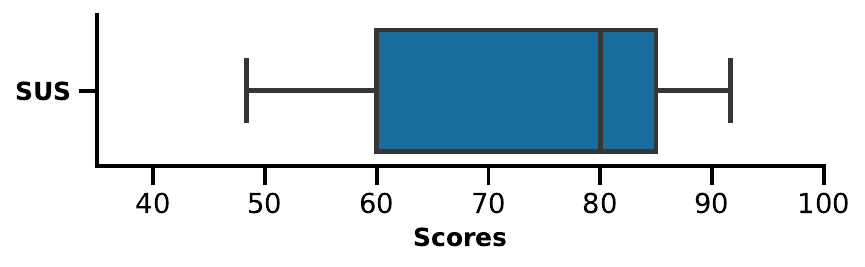}
            \caption{Usability ratings}
            \label{fig:sus}
        \end{subfigure}
        \vspace{0em}
        \begin{subfigure}[b]{\textwidth}
            \centering
            \includegraphics[width=\textwidth]{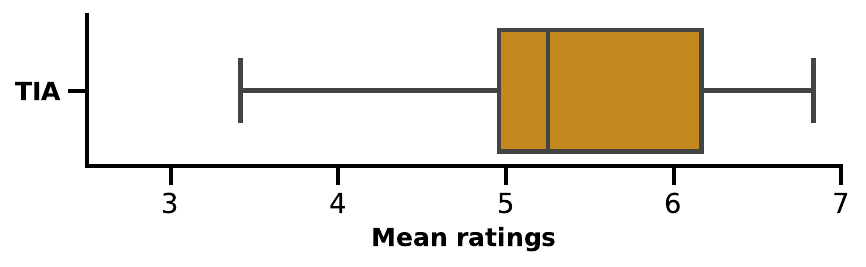}
            \caption{Trust ratings}
            \label{fig:trust}
        \end{subfigure}
    \end{minipage}
    \hfill
    \begin{minipage}[c]{0.49\textwidth}
        \centering
        \begin{subfigure}[b]{\textwidth}
            \centering
            \includegraphics[width=\textwidth]{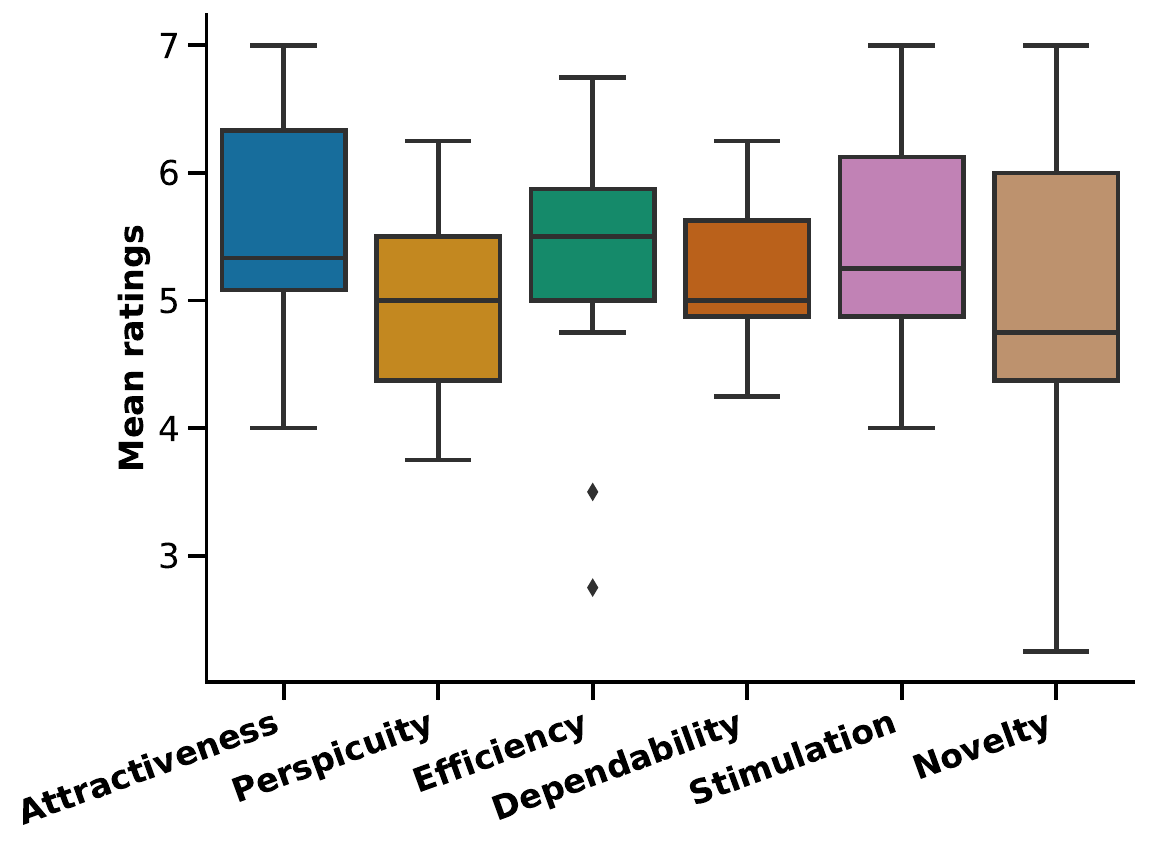}
            \caption{User Experience ratings}
            \label{fig:ueq}
        \end{subfigure}
    \end{minipage}
    \caption{Box plots represent: (a) Users' perceived the toolkit as usable, (b) Users' perceived the toolkit as trustworthy and (c) Users' had positive experience with the toolkit on different UEQ qualities}
    \label{fig:boxplots_all}
\end{figure*}
We conducted a preliminary small-scale user study to evaluate the proposed toolkit. Participants were asked to design three distinct roles/characters for a robot: (1) a hospital receptionist, (2) a mathematics teacher, and (3) a driving assistant. These roles were deliberately selected to include diverse application contexts and to facilitate a comprehensive evaluation of the toolkit. After designing each role, participants were asked to interact with the robot assuming the perspective of an end user. Upon completion of all the tasks, participants filled out a questionnaire hosted on LimeSurvey (\url{https://www.limesurvey.org}). 

The survey collected demographic information, prior technology experience, and responses to several validated instruments that were adapted to fit the context of the toolkit. We assessed the toolkit's usability using the system usability scale (SUS) \cite{SUS}, evaluated trust in the toolkit with the trust in automated systems (TIA) scale \cite{Trust}, and measured user experience using the user experience questionnaire (UEQ) \cite{UEQ}. Participants rated their agreement on a 7-point Likert scale, following best practice recommendations by Schrum et al. \cite{SchrumLikert2023} to capture a wider range of users' perceptions.

A total of 11 participants (9 male, 2 female) were recruited from university students and colleagues. Participants' ages ranged from 24 to 33 years (\textit{M} = 27, \textit{SD} = 2.64), and all held at least an undergraduate degree. Regarding prior experience with AI chatbots, six participants reported frequent (daily) use, four reported moderate (weekly) use, and one reported infrequent use (a few times per year). Additionally, six participants had previous experience with conversational robots. Participants spent on average 30 minutes to complete the study, devoting about 10 minutes to each role design and interaction task. 

Figure \ref{fig:boxplots_all} represents the participant ratings for the SUS, TAS, and UEQ measures. The overall SUS score of 72.87 indicates good level of perceived usability of the toolkit (Figure \ref{fig:boxplots_all}a). UEQ ratings of the toolkit also reflect a generally positive user experience on different qualities (Figure \ref{fig:boxplots_all}c): Attractiveness (\textit{M} = 5.65, \textit{SD} = 0.94), Perspicuity (\textit{M} = 4.90, \textit{SD} = 0.73), Efficiency (\textit{M} = 5.22, \textit{SD} = 1.18), Dependability (\textit{M} = 5.20, \textit{SD} = 0.64), Stimulation (\textit{M} = 5.47, \textit{SD} = 0.89), and Novelty (\textit{M} = 4.90, \textit{SD} = 1.30). Additionally, the toolkit was perceived as trustworthy (Figure \ref{fig:boxplots_all}b) with above average trust ratings (\textit{M} = 5.34, \textit{SD} = 1.08).

\end{sloppypar}

\section{Discussion and Conclusions} 
This study introduced a modular open-source WoZ toolkit for creating customizable social robotic avatars with local LLMs. Our web-based system facilitates multimodal interaction (text, button-activated speech, and wake-word voice command) and gives researchers fine-grained control over robot character/role, language, and interactive design through an intuitive control panel. Our approach integrates local LLMs, enabling more natural and scalable conversational interactions with minimal manual scripting. By using local LLM inference via Ollama, the toolkit ensures low latency and on-device functionality. The ability to rapidly design robots for diverse roles such as hospital receptionists or driving assistants (as seen in our user study) supports broader experimentation and aligns with calls to lower technical barriers for non-expert developers/designers \cite{Smit2025Enhancing}. Users' feedback on the toolkit indicated generally positive ratings across usability, trust, and user experience metrics.

Currently, Google Cloud APIs handle speech-to-text and text-to-speech tasks, creating a dependency on external services. Replacing them with open-source, on-device solutions will ensure full offline operation and stricter data ownership. The user study involved a limited number of participants, most of whom were familiar with chatbot technologies. Broader studies are needed across age groups and levels of technological familiarity to validate usability, user experience, and trust findings. The current version of the toolkit utilizes screen-based avatars to simulate robotic behavior. Conducting real-world evaluations with physically embodied robots similar to QTrobot (\url{https://luxai.com}) could offer valuable insights into embodied interaction, perceived presence, and the system's potential for adaption across diverse HRI applications. 

\begin{sloppypar}
Future research with this toolkit should investigate the impact of specific role types, as well as variations in LLMs, voice characteristics, gender presentation, and avatar appearance. Specifically, the investigation should examine how these design variables impact users’ perceptions, with the objective of determining optimal configurations for various application contexts. Our roadmap focuses on improving and optimizing real-time performance of the toolkit as well as migrating all language and speech processing capabilities to local modules (on-device) to reduce latency and dependency on external services.
The toolkit is reproducible, with the complete source code, required packages, and deployment instructions available at: \url{https://github.com/atikkhannilgar/SRWToolkit}
\end{sloppypar}

\begin{credits}
\subsubsection{\ackname} This work was funded by Honda Research Institute Europe GmbH. We would like to thank all the users who participated in the experiment.

\subsubsection{\discintname}
The authors have no competing interests to declare that are
relevant to the content of this article.
\end{credits}
%
%
%
\bibliographystyle{splncs04}

\bibliography{mybib}

\end{document}